\documentclass[
 reprint,
 amsmath,amssymb,
 aps,
]{revtex4-1}

\usepackage{graphicx}
\usepackage{dcolumn}
\usepackage{bm}

\begin{document}

\preprint{APS/123-QED}

\title{Measurement of  7p$_{1/2}$-state hyperfine structure and 7s$_{1/2}$-7p$_{1/2}$ transition isotope shift in $^{203}$Tl and $^{205}$Tl }
\author{G. Ranjit\footnote{current address: Dept. of Physics, Univ. of Nevada, Reno, NV 89557},  D. Kealhofer\footnote{current address: Dept. of Physics, Univ. of California, Santa Barbara, CA 93106}, G.D. Vukasin, and P.K. Majumder}%
\email{pmajumde@williams.edu}
\affiliation{Department of Physics, Williams College, Williamstown, MA 01267}




\date{\today}

\begin{abstract}
A two-step, two-color laser spectroscopy technique has been used to measure  the hyperfine splitting of the  7p$_{1/2}$ excited state in $^{203}$Tl and $^{205}$Tl, as well as the isotope shift within the 
7s$_{1/2}$ - 7p$_{1/2}$ transition.  Our measured values for the hyperfine splittings, 2153.2(7) MHz (in $^{203}$Tl) and 2173.3(8) MHz (in $^{205}$Tl), each differ by 20 MHz from previously published values which quoted comparable precision. The transition isotope shift of $^{203}$Tl relative to $^{205}$Tl was measured to be 534.4(9) MHz.   In our experiment, one laser was locked to the thallium ground-state 6p$_{1/2}$ - 7s$_{1/2}$  378 nm transition, while the second, spatially overlapping laser was scanned across the 7s$_{1/2}$(F=1) - 7p$_{1/2}$(F=0,1) hyperfine transitions. To facilitate accurate frequency calibration, radio-frequency modulation of the laser was used to create sidebands in the absorption spectrum.
\end{abstract}

\maketitle


\section{\label{sec:level1} Introduction}

Precise measurements of the atomic structure of complex atoms play an essential role in guiding the refinement and testing the accuracy of { \em ab initio} atomic theory calculations.  Accurate approximations for the valence electron wavefunctions of these atoms are a key component in a number of atomic-physics-based tests of elementary particle physics.  In the trivalent thallium atomic system, new calculational techniques make use of a hybrid method combining perturbative features with a configuration interaction approach to address valence electron correlations\cite{Saf08, Saf09}.  In our research program, using both vapor cell and atomic beam spectroscopy techniques, we have completed a series of experimental atomic structure measurements in thallium\cite{Maj99, Richardson00, Doret02},  which are in excellent agreement with these recent calculations\cite{Saf05, Saf06}. When such atomic theory calculations are combined with  an experimental parity nonconservation (PNC) measurement in thallium\cite{Vetter, Kozlov01} the combination provides an important test of standard-model electroweak physics. At present, the current quoted accuracy of the theory lags that of the experiment by roughly a factor of three.  In recent years, a similar calculational approach to that used for thallium has been extended to other Group IIIA systems such as indium\cite{Saf07}.  Very recently, we completed a new precision measurement of atomic polarizability in indium\cite{Ranjit13}, which in turn spurred a new round of wave function calculations\cite{Saf13}.  These results showed excellent agreement at the 1\% level.

In contrast to measurements of transition amplitudes or polarizability, which focus on long-range electron wave function behavior, measurements of hyperfine structure and isotope shifts in these systems can probe short-range wave function models as well as nuclear physics models.  These wave function models are essential in accurately calculating symmetry-violating phenomena in heavy atoms, which are inherently short-range.  Finally, measurements of thallium hyperfine structure (HFS) and isotope shifts (IS), such as reported here, are of direct relevance to recent calculations of the so-called `Schiff' moment in thallium\cite{Porsev12} which are essential for interpreting T-violating electric dipole moment measurements in atomic systems.

In the late 1980s, a group from U. Giessen completed a series of HFS and IS measurements of thallium excited states using pulsed and cw laser excitation\cite{Grexa88, Hermann90}.  A subset of these measurements were later corrected due to self-reported `calibration and linearization errors'\cite{Hermann93}.  More recently, two measurements of the 7s$_{1/2}$-state HFS, first by our group in 2000\cite{Richardson00}, and then by Chen \emph{et al.} in 2012\cite{Chen12} have confirmed the inaccuracy of the original set of measurements from the Giessen group.  In the present paper, we report new measurements of the 7p$_{1/2}$ excited-state HFS in the two naturally occurring isotopes of thallium.  While results for these intervals were reported by the Giessen group in 1988\cite{Grexa88}, those particular intervals were never re-measured.  As discussed below, our new results for both isotopes are roughly 20 MHz larger than those reported in \cite{Grexa88}.  Given the $ \sim$ 1 MHz quoted uncertainties of both sets of results, it is possible that the calibration errors in the older results can again explain this discrepancy.  In an effort to pay particular care to frequency-axis linearization and calibration, we have employed both a stable Fabry-Perot cavity, as well as an electro-optic modulator (EOM), which provides a set of spectral sidebands at precisely known frequency separations in our atomic spectra.   Looking ahead, there is strong motivation to re-measure a number of excited state hyperfine splittings in thallium, as it is essential to provide accurate experimental benchmarks for ongoing \emph{ab initio} atomic theory calculations.
  
In the past few years, the development of GaN semiconductors has led to the production of laser diode radiation in the blue and near-UV wavelength range. We have made use of such a laser source, operating at 377.6 nm, in conjunction with a second InGaAs diode laser operating at 1301 nm,  to excite thallium atoms in a heated quartz vapor cell to the 7p$_{1/2}$ state.   This two-step excitation scheme, as well as the modulation / lock-in detection method outlined below, provides Doppler-free, zero-background signals when we detect the infrared laser transmission through the thallium cell. The availability of two relatively inexpensive, low-power diode laser systems makes this method a practical alternative to a two-photon excitation scheme using a higher-power red laser.

\section{Experimental Details}
\subsection{Spectroscopy scheme and laser locking}

\begin{figure}[t]
\includegraphics[scale = 0.21]{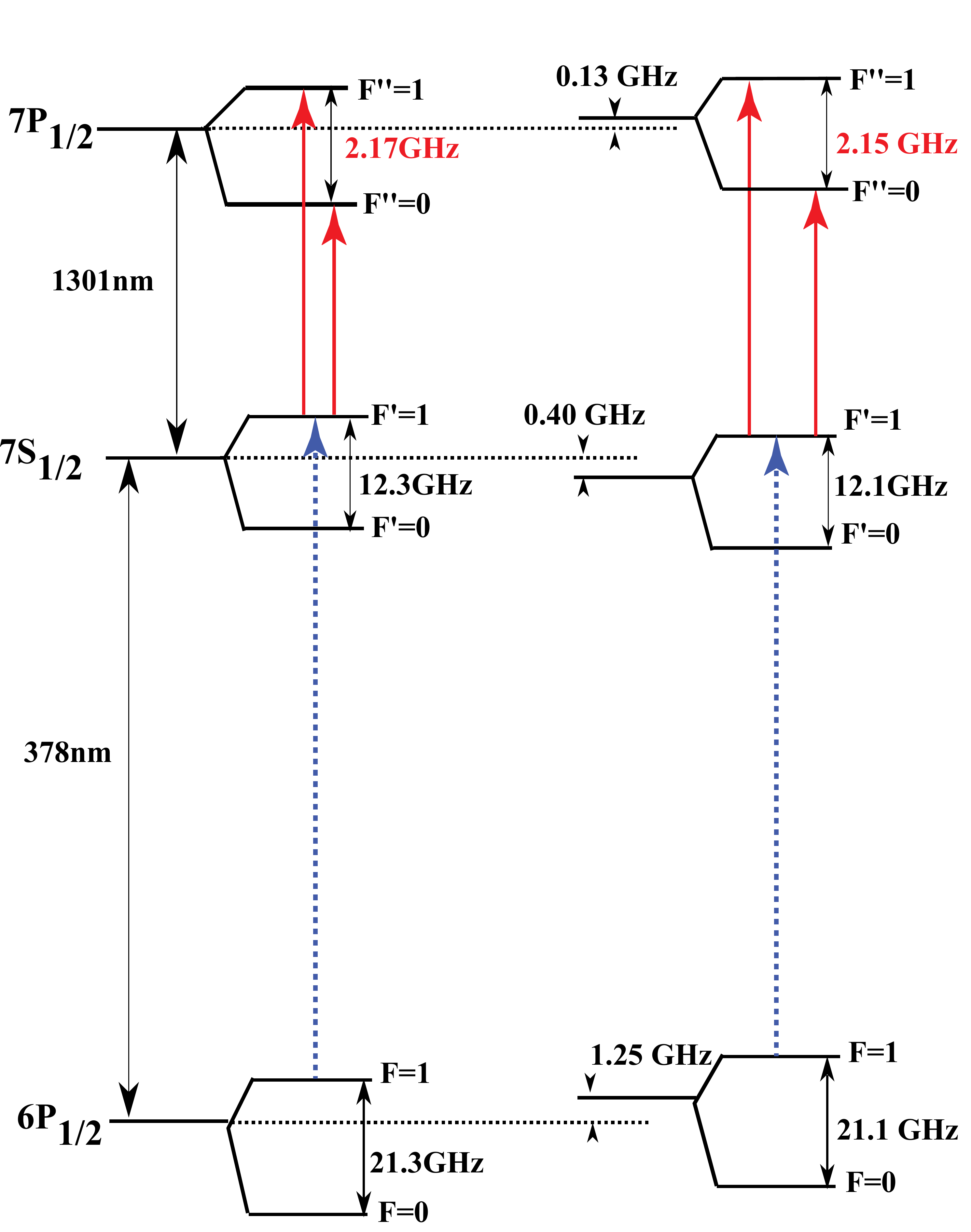}
\caption{\label{thalliumenergylevels}(Color online) A partial energy level diagram for  $^{205}$Tl (left) and $^{203}$Tl (right). The 377.6 nm laser is locked to the first-step transition (dashed blue line), exciting atoms to the intermediate state. The 1301 nm laser is then scanned across the second-step transitions (solid red arrows) to produce hyperfine spectra for the 7p$_{1/2}$-state.}
\end{figure}

Fig. \ref{thalliumenergylevels} shows the relevant energy levels of the two naturally-occuring thallium isotopes:  $^{205}$Tl (70.5\% abundance) and $^{203}$Tl (29.5\% abundance).  Both isotopes have nuclear spin $I=1/2$.  In order to probe the 7p$_{1/2}$-state hyperfine structure, we begin by locking a UV laser to the  6p$_{1/2}$(F=1) $-$ 7s$_{1/2}$(F$^\prime$=1) ground-state transition at 377.60 nm.  Stabilization of this first-step laser is essential to minimize drift and instability in the resonance frequency of the second-step transition. A second infrared (IR) laser is then scanned across the the 7s$_{1/2}$(F$^\prime$=1) $-$ 7p$_{1/2}$(F$^{\prime\prime}$=0,1) hyperfine transitions. Since the first-step transition features a large (1.6 GHz) isotope shift,  the UV absorption spectrum shows partially resolved isotopic peaks, even in our Doppler-broadened vapor cell environment.  We are able to lock the UV laser to either the center of the $^{205}$Tl resonance, the center of the $^{203}$Tl resonance, or to a point between the isotopic resonances where, due to the Doppler broadening, non-zero velocity classes of each isotope (one blue-shifted, the other red-shifted) can be simultaneously excited.  The former lock positions are exploited to probe the 7p$_{1/2}$ HFS of each thallium isotope separately, while the latter dual-isotope (DI) lock point is used to extract the isotope shift, once the relative Doppler shifts of the isotopes are removed (as described below).  

\begin{figure}
\includegraphics[scale = 0.35]{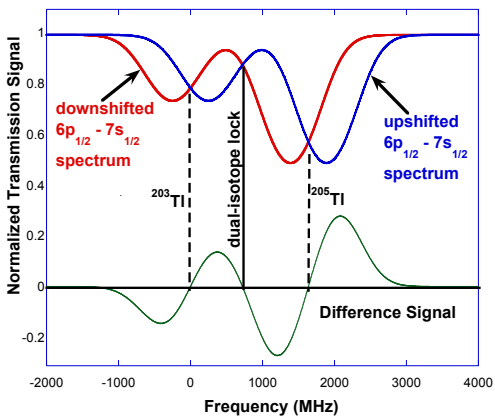}
\caption{\label{diffsig}(Color online) Thallium 6p$_{1/2}$(F=1) $-$ 7s$_{1/2}$(F$^\prime$=1) vapor cell transmission spectra resulting from AOM down-shifted (red) and up-shifted (blue) laser beam components.  The lock points of the UV laser for $^{203}$Tl, dual-isotope, and $^{205}$Tl excitation are indicated, and the difference signal used for laser locking is shown below.}
\end{figure}

The locking scheme is based on a method developed in our group\cite{Gunawardena08, Gunawardena09}.  Briefly, we pass the UV laser beam through an acousto-optic modulator (AOM), producing diffracted beams which are frequency shifted by  $\pm$250 MHz from the incident beam.  These diffracted beams are polarized orthogonally using half-wave plates, combined in a polarizing beam splitter (PBS) and sent through a small, supplementary oven containing a thallium vapor cell.  For our cell, an oven temperature near 500 $^\circ$C produces roughly 50\% absorption, ideal for our locking procedure.  The frequency-shifted beams are separated after transmission through the cell, and directed into a differential photodetector.  This difference signal is the input to a standard servo circuit which steers the laser frequency via a voltage applied to the piezoelectric transducer (PZT) controlling the diffraction grating in the laser cavity.  We find that this locking technique reduces the drift of the laser frequency to 1 MHz or less over time scales of several hours\cite{Gunawardena08}.  Fig. \ref{diffsig} shows the difference signal produced from the frequency-shifted transmission spectra of the relevant thallium 6p$_{1/2} $ $-$ 7s$_{1/2}$ hyperfine transition.  Indicated at the difference signal zero crossings are the three lock points mentioned above.  

When we lock the UV laser to one of the two `single-isotope' locations, the exact lock point along the linear portion of the difference curve is not critical. While the exact lock point determines the velocity class selected for the single-isotope, and thus the exact resonance frequency of the second-step transition, it has no effect on measured frequency differences, such as the hyperfine splitting value itself.  Also, while the non-targeted isotope is in this case significantly off resonance, a very small amount of this secondary isotope, highly Doppler-shifted, will nevertheless be excited.  In our spectra we do not see visible evidence for this contamination at our level of statistical sensitivity.  To ensure that no systematic error results from the presence of this secondary component in the single-isotope spectra, we collect data with UV and IR laser beams in a co-propagating geometry (CO) , as well as a counter-propagating (CTR) geometry.  These configurations should produce identical spectra for the target isotope, while reversing the relative Doppler shift of the spurious second isotopic component.  Consistency in our results upon this reversal (as we observe) gives us confidence that we are free of this potential systematic error.

In the case of dual-isotope excitation, the precise lock point of the UV laser determines the magnitudes of the relative red and blue Doppler shifts for the two isotopes.  Regardless, we know that the sum of magnitudes of these shifts must equal the observed isotopic separation in the 6p$_{1/2}$(F=1) $-$ 7s$_{1/2}$(F$^\prime$=1) hyperfine transition to which we lock.  This separation depends both on the transition isotope shift and the hyperfine anomalies of the 6p$_{1/2}$ and 7s$_{1/2}$-states, all of which have been precisely measured previously\cite{Richardson00, Chen12} yielding the value 1636(1) MHz for this transition.  The eventual isotopic positions in our second-step DI spectra thus reflect both the true isotopic shift of the infrared transition, as well as the relative Doppler shift imprinted by the UV laser lock.  Since the observed Doppler shift is proportional to the laser frequency, the total relative Doppler shift eventually observed in the IR spectra is given by $(f_{IR}/f_{UV}) \times 1636$ MHz = 474.7(4) MHz.  As discussed below, our DI spectra consist of CO and CTR geometries obtained \emph{simultaneously}.   By taking appropriate averages and differences in observed peak splittings, we can use our DI spectra to extract the true 7s$_{1/2}$ $-$ 7p$_{1/2}$ transition isotope shift (TIS) as well as the HFS for both isotopes.  At the same time, as a check, we can confirm that the sum of the isotopic Doppler shifts agrees with the predicted value.

\begin{figure}
\includegraphics[scale = 0.25]{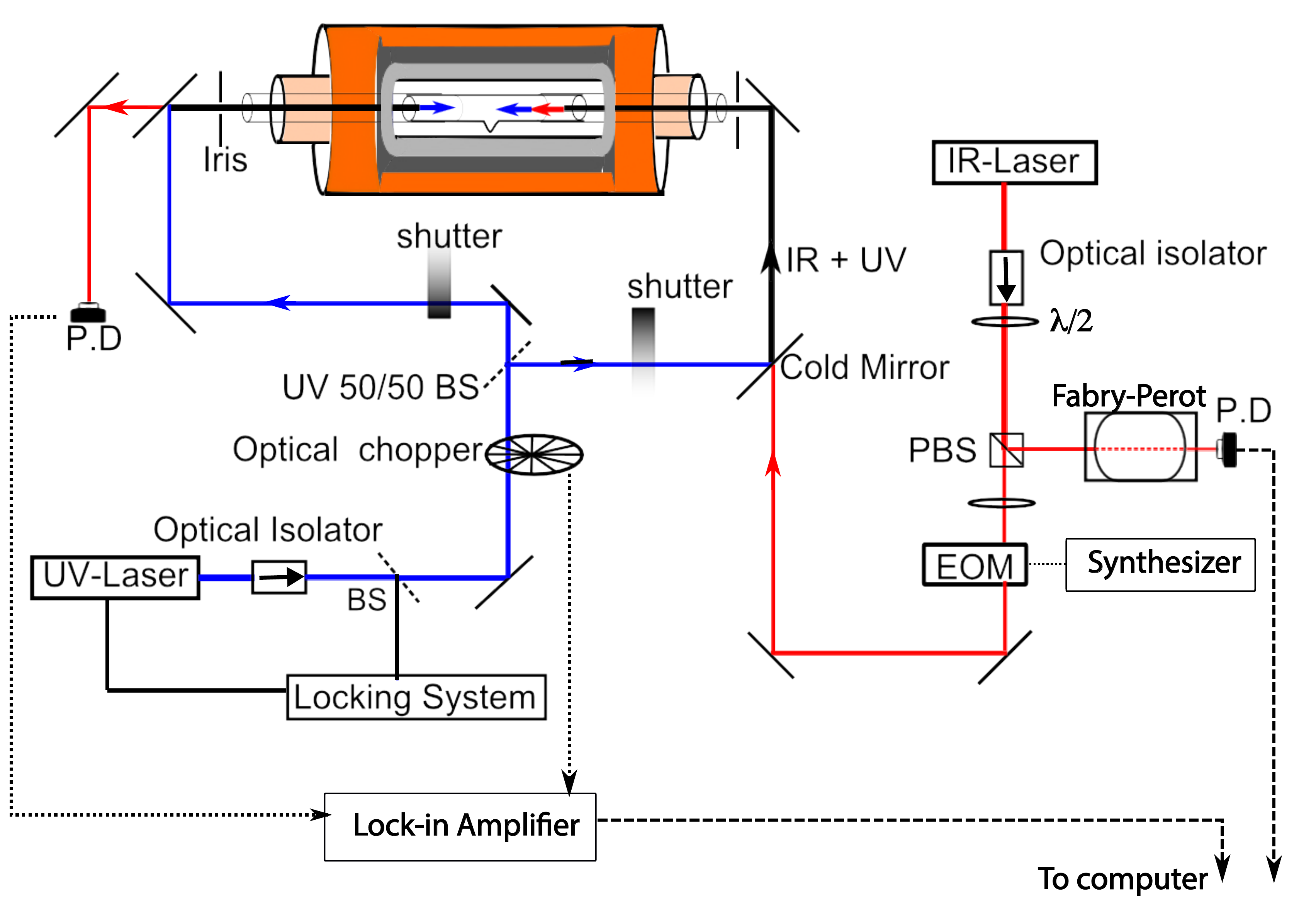}
\caption{\label{exptsetup}(Color online) Experimental layout showing optical setup and interaction region for the two-color spectroscopy scheme.  The cutout shows the UV beam in both CO and CTR configuration along with the IR beam as they pass through the heated vapor cell.  }
\end{figure}

\subsection{Optical System and Experimental Layout}

Fig. \ref{exptsetup} shows a schematic of the two-color spectroscopy apparatus. The experiment uses two commercial external cavity diode lasers (ECDL) in Littrow configuration (both Sacher Lasertechnik, Lynx model). The UV laser driving the first-step transition produces light at 377.6 nm and is locked to the 6p$_{1/2} $( F = 1) $-$ 7s$_{1/2}$ (F$^\prime$ = 1) hyperfine transition, as described above.  While the $\pm 1$-order beams from the AOM are used for laser locking, the undiffracted zero-order beam, whose frequency is located precisely at the lock point of interest, is sent to the interaction region.  This beam is split into two components directed in both co- and counter-propagating directions with respect to the IR beam as they pass through the vapor cell.  We use 1-mm-diameter apertures  placed roughly 1 meter apart on either side of the experimental oven to guide our laser beams, aid in alignment, and ensure that any residual non-parallelism of the beams results in negligible Doppler-shift-related systematic errors, as discussed below.  The second ECDL, tuned to the 1301.3 nm 7s$_{1/2}$(F$^\prime$=1) $-$ 7p$_{1/2}$(F$^{\prime\prime}$=0,1) transitions in thallium, is  steered in a single direction through the same guiding apertures and overlaps both UV laser beams in the heated cell.  The transmitted IR beam is  directed into an InGaAs photodiode detector.  We combine and separate the UV and IR laser beams via dichroic mirrors on either side of the interaction region.  We work to make the diameters and divergences of all three laser beam components as similar as possible.  Prior to entering the oven, a portion of the IR beam is sent to a confocal Fabry-Perot (FP) cavity for frequency linearization and calibration.  The cavity (Burleigh RC-110) has a finesse of roughly 50, and a free spectral range (FSR) near 500 MHz.  It is constructed out of low-expansion material and is contained in an insulated box for passive thermal stability.  The transmitted spectrum of the FP cavity is collected by our data acquisition system as we scan the IR laser.  As an independent calibration method, we pass the IR laser beam through an electro-optic modulator (New Focus model 4423).  The EOM is driven by a 600 MHz synthesizer.  The RF power is adjusted to produce first-order sidebands in the laser spectrum, while keeping higher-order components negligible.  We note that we do not perform FM spectroscopy here (which would involve high-frequency demodulation of the transmitted signal).  Instead we simply detect the IR transmission signal with its additional spectral peaks at precisely known frequency separations.

Computer-controlled shutters block and unblock the UV beam propagating parallel (CO) or antiparallel (CTR) to the IR beam as desired.  As mentioned above, for the single-isotope measurements, we alternate scans in the two geometries, noting that the measured HFS should not depend on this beam geometry.  For the dual-isotope measurements, both shutters are kept open.  

Because the UV laser only interacts with a small fraction of the Doppler-broadened ground state atoms (roughly the ratio of the homogeneous to inhomogeneous line widths, or $\sim$3\%), we expect the direct IR absorption signal to be very small.  To address this, we use an optical chopping wheel to modulate the UV beam as it exits the locking setup (see Fig. \ref{exptsetup}) and thus modulate the population of the intermediate 7s$_{1/2}$ state (at a frequency of $\sim$1.5 kHz). We detect the transmitted IR signal in a lock-in amplifier (SRS model 810) using the chopping frequency as the reference. The lock-in technique thus provides zero-background, high signal-to-noise ratio atomic hyperfine spectra (see figures below).

\subsection{Oven assembly and interaction region}
The  experimental oven consists of  two pairs of clamshell heaters inside a thick insulating layer of fiberglass, placed inside a 1-m-long cylindrical $\mu$-metal housing.  The $\mu$-metal reduces transverse magnetic field components by two orders of magnitude.  The frame is wrapped with a solenoid to cancel the longitudinal component of the earth's magnetic field, so that all components of the field in the interaction region are less than 1$\mu$T.  Within the clamshell heaters, the interaction region consists of a 1-m-long,  6-cm-diameter alumina tube that houses the 10-cm-long, 2.5-cm-diameter thallium vapor cell and ensures uniform distribution of the heat across the cell. The pairs of heaters are slightly separated, the gap allowing the central cell stem containing thallium metal to remain slightly cooler than the cell window faces. A pair of empty quartz tubes, with diameter matching that of the cell, lie flush with the faces of the cell.  These tubes protrude well outside of the hot oven region, minimizing convective air currents near the cell faces.  A thermocouple is placed inside the interaction region to monitor the temperature of the vapor cell. A temperature controller regulates the temperature of the cell in the 400-450 $^\circ$C range via solid state switches which control the duty cycle of the applied AC current. 

\subsection{Data acquisition and experiment control}

Having positioned the UV laser at the desired single- or dual-isotope lock point, we carefully align the co- and counter-propagating UV beams and the IR beam to achieve maximum overlap.   We find that misalignment decreases the IR absorption signal, and, more important, can result in line shape asymmetry and thus potential systematic errors in our determination of frequency splittings.  The data acquisition program sets the appropriate optical shutter configuration for the UV laser beams and initiates a stepwise sweep of of the IR laser upward and then downward in frequency over a 5 - 7 GHz range centered on the 7s$_{1/2}$ $-$ 7p$_{1/2}$ state hyperfine transitions. We sweep the laser frequency by applying a voltage ramp to the PZT which controls the diffraction grating of the ECDL.  A LabVIEW program samples both the lock-in amplifier output signal as well as the FP cavity transmission signal, collecting roughly 1000 data points over the 4 s duration of the laser scan.  Data from up and down sweeps are stored separately for later analysis.  We collect single-isotope data alternately in CO and CTR configurations, opening and closing shutters after each complete laser sweep. Dual-isotope data sets are collected with both shutters open.  In the DI case, due to the complexity of the eight-peak spectrum, we turn off the EOM, removing the FM sidebands.    An individual data set  consists of several hundred up/down laser scans obtained under nominally identical conditions over the course of several hours.  Between data sets, we realign the optical beams and change experimental parameters such as laser sweep speed and extent, UV and IR laser power, relative polarization of the lasers, and the oven temperature.  

\section{Data analysis and results}

\subsection{Linearization and calibration of scans}
The data analysis procedure begins with linearization of the frequency scale via analysis of the Fabry-Perot transmission spectrum.  By insisting that the FP peaks are equally spaced in frequency, we can remove small but reproducible nonlinearities in the frequency sweep due to the non-linear and hysteretic response of the PZT to applied voltage.  Specifically,  we fit our FP spectrum to an Airy function in which the frequency argument is expressed as a fourth-order polynomial of the point number. We find that higher-order polynomials do not improve the statistical quality of the fit.  

This procedure linearizes the frequency axis, but does not address absolute calibration. For all of our linearized scans,  the working calibration is based on the estimated FP FSR of 500 MHz.  For our ultimate calibration, we make use of the FM sidebands from our EOM, as described in the next section. However, as an important cross-check on that method, we performed the following direct determination of the FSR of our Fabry-Perot cavity.  With the use of a 0.1 ppm wave meter (Burleigh WA-1500; 30 MHz resolution), we tuned our laser to a particular transmission peak of the FP cavity and recorded the optical frequency.  We then repeated this procedure for many individual frequencies over a roughly 100 GHz range, ensuring in each case that the laser was operating in a single-mode fashion.  Starting with a good estimate of the FSR of the cavity, we considered all difference pairs of optical frequencies, which we require to be an integer multiple of the FSR.  Choosing the best fit integer in each case gave us many independent determinations of the FSR, and the average of these produced an FSR value of increased precision.   This procedure was repeated several times over the course of one month, and we found that the average measured FSR value was 501.2(3) MHz.  Since all of the linearized frequency scales for our atomic spectra are based on the nominal FSR value of 500.0 MHz, we define a frequency-scale correction factor which can be applied to all measured frequency intervals, $\mathcal{C}_{FP} = 501.2/500.0 = 1.0024(5)$.  We can compare this result to the FM-sideband-based value discussed next.

\subsection{Results from single-isotope spectra}

\begin{figure}[t]
\includegraphics[scale = 0.43]{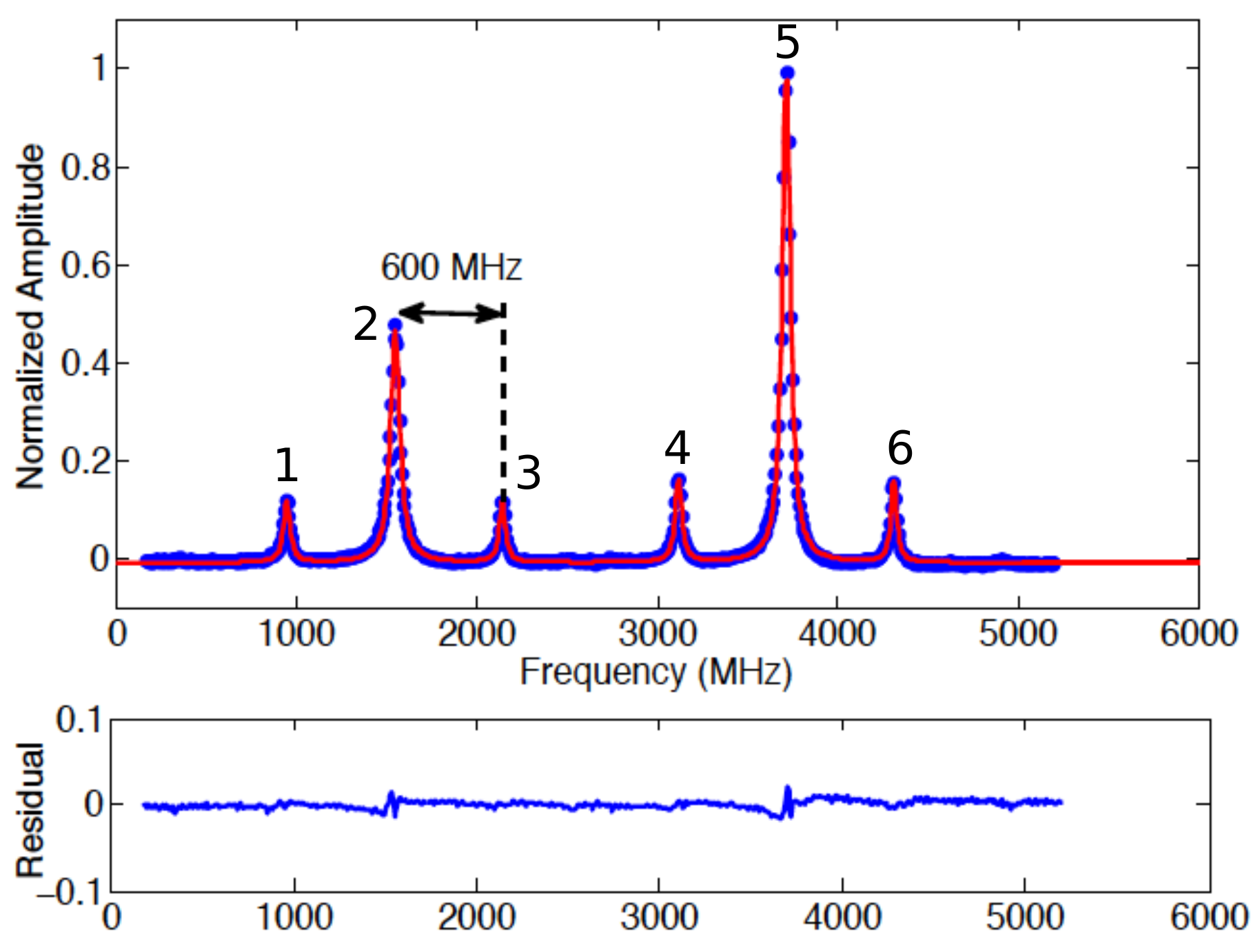}
\caption{\label{600MHz} (Color online) A typical 4 s scan across  the 1301 nm 7s$_{1/2}$ (F $'$= 1) $\rightarrow$ 7p$_{1/2}$ (F$''$ = 0, 1) hyperfine transitions of the $^{203}$Tl isotope. The 600 MHz first-order FM sidebands are clearly visible. Blue data points as well as a red fitted curve (solid line) are shown, with corresponding (expanded) residuals for the fit shown below. }
\end{figure}

The next phase of our analysis for either single-isotope $^{203}$Tl or $^{205}$Tl spectra is to determine the 7p$_{1/2}$-state hyperfine splitting with a properly calibrated frequency scale.  As the FM sidebands used for frequency calibration are a built-in part of the hyperfine spectra, we begin by discussion the overall line shape analysis procedure.  Given the two-step nature of our excitation scheme, in which the UV laser selects a single velocity class of the vapor cell atoms, our IR spectra should be inherently Doppler-free.  In practice, non-zero divergence of the overlapping laser beams leads to some residual Doppler broadening, but it is small compared to the (power-broadened) homogeneous line width ($\sim$50 MHz).  We have analyzed the atomic spectral peaks in terms of Voigt convolution profiles to incorporate the Gaussian component width, but find negligible difference in the quality of fits, and no measurable change in determination of peak locations, when we instead employ a simpler fit to a sum of Lorentzians.  These Lorentzians are assigned a single common fit parameter for the spectral width, while individual line centers and amplitude parameters are determined. As shown in Fig. \ref{600MHz}, we obtain a six-peak spectrum including the F= 0 and 1 hyperfine peaks (labeled `2' and `5' respectively in  Fig. \ref{600MHz}) as well as first order sidebands for each, located at precisely $\pm$ 600 MHz from the main peaks.  Spectra for the other ($^{205}$Tl) isotope look identical except for a roughly 20 MHz larger hyperfine splitting.  As mentioned, good laser beam overlap minimizes spectral peak asymmetry, though in principle, a common asymmetry in all peaks should not affect peak separation measurements.  While we find that very small residual asymmetries persist (note expanded residuals at the bottom of Fig. \ref{600MHz}), these are effectively randomized by re-alignment of beams over the course of many distinct  data sets.  An individual scan such as shown in Fig. \ref{600MHz} yields statistical uncertainties in peak positions of several MHz.  We sorted our fit results for each isotope into sub-categories by laser sweep direction and laser beam geometry (CO vs. CTR) for further study of potential systematic errors.  In the discussion below, we define frequency splittings $\delta\nu_{ij}$ to refer to the $i - j$ peak frequency difference.

\begin{figure}[t]
 \includegraphics[scale = 0.43]{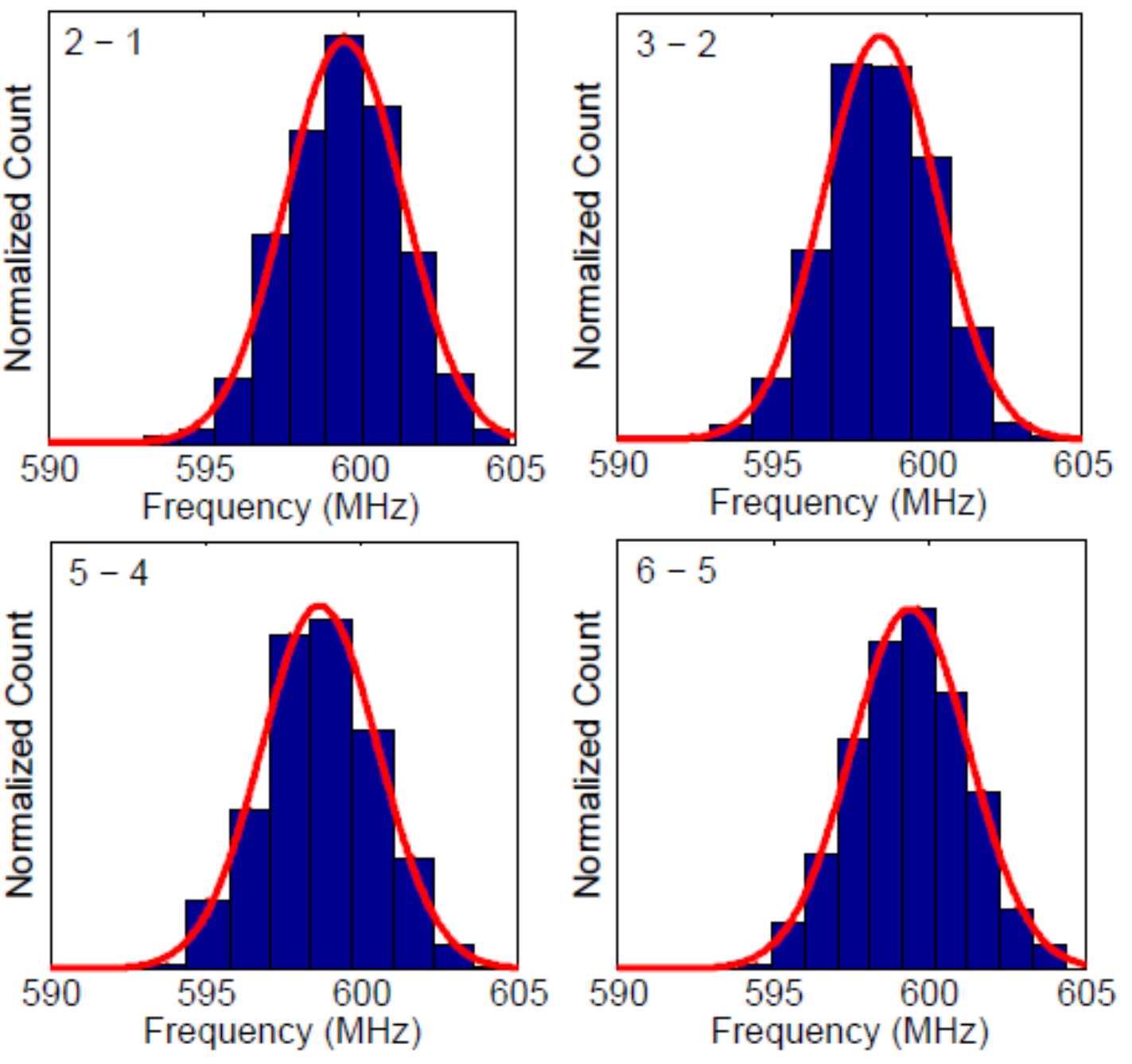}
 \caption{\label{4sidebands} (Color online) For one day's worth of data, we show histograms of each sideband splitting as measured from the fitted six-peak experimental spectrum.  Solid lines indicate fitted Gaussians.  The average of the central value of these histograms provides the calibration factor to correct the frequency scale, while differences among peak values allows us to study residual frequency-scale non-linearity.}
\end{figure}

Having completed fits and located all peaks using our nominal frequency axis, we considered the four sideband splitting values by computing differences in relevant peak positions (specifically, $\delta\nu_{21}$, $\delta\nu_{32}$, $\delta\nu_{54}$, and $\delta\nu_{65}$).  Because the true FP cavity FSR is greater than our nominal 500 MHz value, we expect these measured intervals in our spectra to deviate slightly from 600.0 MHz.  We study both the consistency in these frequency values among the four splittings, and we also compare the results of this calibration method to the direct FSR measurement approach.   Fig. \ref{4sidebands} shows a histogram of the four sideband splitting values derived from several hundred individual scans taken on a single day.  We find that the average value of these four measured sideband splittings is 598.8(1) MHz for this particular set of data.  An analogous calibration factor can be extracted using this method by simply dividing this value into the `known' sideband splitting value of 600.0 MHz.  After analyzing all spectra taken under varying experimental conditions for both isotopes, we find that the average calibration factors for all HFS data is $\mathcal{C}_{EOM} = 1.0020(2)$, in good agreement with the independent FP calibration value quoted above.  As discussed below, studying systematic differences among the four sideband splitting values allows us to assess the effectiveness of our frequency scan linearization procedure, and place systematic error limits on how such nonlinearity would affect our calibration values.  It is a simple matter, finally, to multiply the raw value of the measured HFS splitting by the appropriate correction factor to produce final HFS values.  For example, the hyperfine splitting corrected via the FM-sideband calibration would be given by $\delta\nu^{\mathcal{(C)}}_{52}\equiv \mathcal{C}_{EOM} * \delta\nu_{52}$.

Our spectral analysis provides a second, independent method of calculating HFS.  Looking again at our six peak analysis and Fig. \ref{600MHz}, we note that the 4-3 peak frequency difference ($\delta\nu_{43}$) , when added to exactly twice the sideband splitting value, should also yield a value of the HFS.  In fact, since the actual measured frequency splitting here is less than half of the total HFS, this method of obtaining the hyperfine splitting is less sensitive to potential calibration errors, which scale directly with the magnitude of the frequency splitting.   Though the sideband peaks are smaller in amplitude, we find that the signal-to-noise ratio of these peaks is sufficient to provide a precise, statistically-independent measurement of the hyperfine splitting. 

\subsection{Results from dual-isotope spectra}

The first phase of analysis for our dual-isotope spectra mirrors  the single-isotope analysis procedure exactly:  we use the Fabry-Perot transmission data to create a linearized frequency scale with our nominal frequency axis scaling.  Since we do not utilize the EOM for these data scans, we must eventually apply one of the calibration correction factors discussed above as a final step in determining frequency intervals.  

\begin{figure}
 \includegraphics[scale = 0.30]{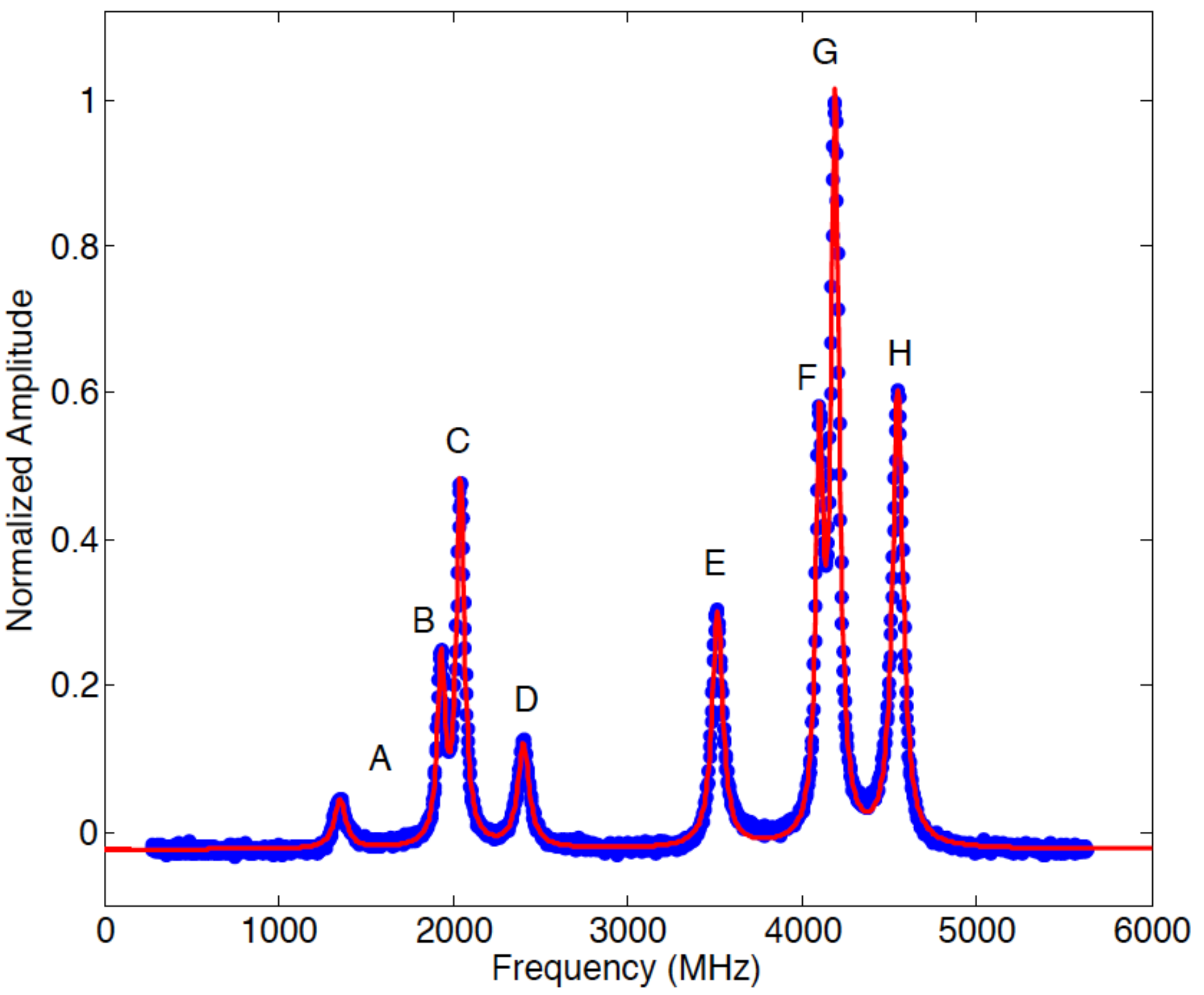}
\caption{\label{dual}(Color online) The eight peak spectrum corresponding to both CO and CTR geometries  when we excite both isotopes and scan across transitions to both the F=0 and F=1 hyperfine levels of the 7p$_{1/2}$ level. The fit to a sum of eight Lorentzians (red solid line) is superimposed on the blue data points.}
\end{figure}

As discussed in section II A, in order to gain information regarding the isotope shift in this 1301 nm transition, we must simultaneously excite both isotopes.  Given that we use a single UV laser frequency, our dual-isotope excitation lock point is red-detuned from the $^{205}$Tl resonance and blue-detuned from the$^{203}$Tl resonance.  We thus excite $^{205}$Tl atoms which are moving with non-zero longitudinal velocity towards the UV beam, and the reverse for the $^{203}$Tl atoms.  Since the UV laser is sent bi-directionally through our vapor cell, we excite two classes of moving atoms for each isotope.    The second-step IR laser will then excite hyperfine transitions that reflect not only true isotopic shifts, but also these relative Doppler shifts.  Because the IR laser frequency is roughly 1/3 that of the UV laser, these Doppler shifts are substantially smaller than the original isotopic shift in the ground-state UV transition.  As it turns out, given the sign of the isotope shift, the relative Doppler shift measured in the CO laser beam geometry nearly doubles the true isotope shift, while that from the CTR geometry tends to nearly cancel it.  We can average the appropriate peak splittings measured for the two geometries to obtain the desired quantity (the potential systematic error resulting from non-parallel beams is discussed in Sec. IV).   Figure \ref{dual} shows the eight-peak dual-isotope spectrum.  The first (last) four peaks represent transitions to the 7p$_{1/2}$ F=0 (F=1) hyperfine level.  Peaks labelled A, B, E, and F originate from the less abundant $^{203}$Tl isotope, whereas peaks C, D, G, and H derive from $^{205}$Tl.  

Defining the frequency $f_0$ to be the $^{205}$Tl 7s$_{1/2}$-7p$_{1/2}$ optical transition frequency in the absence of hyperfine structure (\emph{i.e.} the center of gravity of these transitions), we can parametrize each of the eight peak frequencies in terms of $\mathcal{H}$, the 7s$_{1/2}$ and 7p$_{1/2}$-state hyperfine splittings, $ I_{7s-7p}$, the transition isotope shift of $^{203}$Tl relative to $^{205}$Tl, and $\Delta f_{(IR)}$, the magnitude of the IR Doppler shift for each isotope.  We find for the four peaks in Fig. \ref{dual} corresponding to the co-propagating configuration:
\begin{eqnarray}
\nu_A &=& f_0 - \frac{3}{4} \mathcal{H}_{7p,205} - \frac{1}{4} \mathcal{H}_{7s,205}   - \Delta f_{205(IR)} \nonumber\\
\nu_D &=& f_0 - \frac{3}{4} \mathcal{H}_{7p,203} - \frac{1}{4} \mathcal{H}_{7s,203}   + I_{7s-7p} + \Delta f_{203(IR)} \nonumber\\
\nu_E &= &f_0 + \frac{1}{4} \mathcal{H}_{7p,205} - \frac{1}{4} \mathcal{H}_{7s,205}   - \Delta f_{205(IR)} \nonumber\\
\nu_H &=& f_0 + \frac{1}{4} \mathcal{H}_{7p,203} - \frac{1}{4} \mathcal{H}_{7s,203}   + I_{7s-7p} + \Delta f_{203(IR)}. \nonumber
 \end{eqnarray}
The corresponding counter-propagating peaks ( B, F, C, and G in Fig. \ref{dual}), have nearly identical parametrization to those above with the exception of a sign reversal in the final term in each case.  The presence of the $\mathcal{H}_{7s}$ terms in each expression reflects the fact that all transitions in this spectrum originate from the F=1 hyperfine level of the 7s$_{1/2}$ state.  It is important to note that differences in isotopic hyperfine splitting for both 7s$_{1/2}$ and 7p$_{1/2}$ levels (the hyperfine anomaly) must be accounted for in order to isolate the isotope shift, since the latter quantity is defined to be the frequency shift which would be measured in the \emph{absence} of hyperfine structure.  Various combinations of the peak frequencies can be constructed to extract all of the frequency splittings of interest.   For example,  the $^{205}$Tl hyperfine splitting in the 7p$_{1/2}$ state can be found by taking the difference between peaks E and A ($\delta\nu_{EA}$) or peaks F and B ($\delta\nu_{FB}$).  When these HFS intervals are calibration-corrected, we can compare their value to those derived from our single-isotope experiments.

To calculate the transition isotope shift, we take appropriate averages of CO and CTR peaks to remove the Doppler shift, then remove the hyperfine anomalies in both the 7s$_{1/2}$ and 7p$_{1/2}$ states.  Thus:
\begin{equation}
\begin{split}
I_{7s-7p} &=\frac{3 \bigg\vert{\frac{\nu_G + \nu_H}{2}-\frac{\nu_E+\nu_F}{2} }\bigg\vert  + \bigg\vert{\frac{\nu_C + \nu_D}{2}-\frac{\nu_A+\nu_B}{2} }\bigg\vert}{4}\\
& + \frac{\mathcal{H}_{7s,203} - \mathcal{H}_{7s,205}}{4}, 
\label{isotopeshift}
\end{split}
\end{equation}
where $\mathcal{H}_{7s,203} $ and $\mathcal{H}_{7s,205}$ are the previously measured hyperfine splittings in the 7s$_{1/2}$ states in $^{203}$Tl and $^{205}$Tl isotopes respectively\cite{Richardson00, Chen12}.   The small uncertainty in these measured 7s$_{1/2}$ HFS splittings does not significantly increase the overall measurement error in our isotope shift determination.  In addition to the isotope shift, we can independently extract  7p$_{1/2}$ HFS values for each isotope from analysis of the dual-isotope spectra.  In some cases we discovered residual peak asymmetry due to slight misalignment of either the CO or the CTR UV beam, leading to poorer quality fits and  HFS values which were not in good agreement with the single-isotope results.  These scans were discarded. The majority of the the dual-isotope scans, however, showed good consistency in HFS values when compared to the single-isotope results.  In our final analysis, we used roughly 1000 individual dual-isotope scans.  From data set to data set, these results showed somewhat more scatter than was observed for the case of single-isotope data, likely due to the tendency for greater line shape asymmetry associated with the challenging of overlapping three laser beams.  Nevertheless, the process of realigning the beams over the course of one day, and between days, did randomize this variation, as could be observed by considering the overall distribution of our results.  Our final statistical uncertainty in the isotope shift value reflects the observed data set to data set variation.

Finally, it is easy to see that various differences in peak frequencies yield a quantity equal to the sum of the blue and red Doppler shifts of the two isotopes in the IR spectrum.  We can average all of those differences to obtain
\begin{eqnarray}
\label{Doppshift}
| \Delta f_{205(IR)}| +| \Delta f_{203(IR)}| =  \nonumber \\  
 \frac{1}{4} [\delta\nu_{BA} + \delta\nu_{DC} + \delta\nu_{FE} +\delta \nu_{HG}]. & &
\end{eqnarray}
As discussed in Sec II, regardless of the precise UV laser lock point in the region between the isotopic resonances, the sum of the Doppler shifts which we observe in the IR spectrum should reflect this UV transition isotopic separation.  As an additional check on the consistency of our overall measurement scheme, we can compare the measured peak frequency combination such as the RHS of the equation with the expected value near 475 MHz.

\section{ Exploration of Systematic Errors}

We explored a variety of potential systematic errors in our experiment using several different methods.  Over the course of roughly one month of data collection, we varied the IR laser power over a factor of five in a series of steps and found no statistically resolved change in the measured hyperfine splittings.  We also varied the relative polarization of the UV and IR lasers used for our two-step excitation.  We saw significant changes in relative  peak \emph{heights}, but no measurable change in frequency splitting.  We also explored the effect of varying the thallium vapor cell temperature.  At very high thallium vapor pressures (in our case, for cell temperatures above  550$^\circ$C), we saw evidence in our spectra of Doppler-broadened `pedestals', which we attributed to radiation trapping effects at these high densities.  These spectral features disappeared at lower temperatures.   Our typical operating temperature for the data presented here was 400$^\circ$C, but we observed no visible difference in our spectra, and our results were consistent at the 0.2 MHz level when we operated at 450$^\circ$C.

\begin{figure}
 \includegraphics[scale = 0.40]{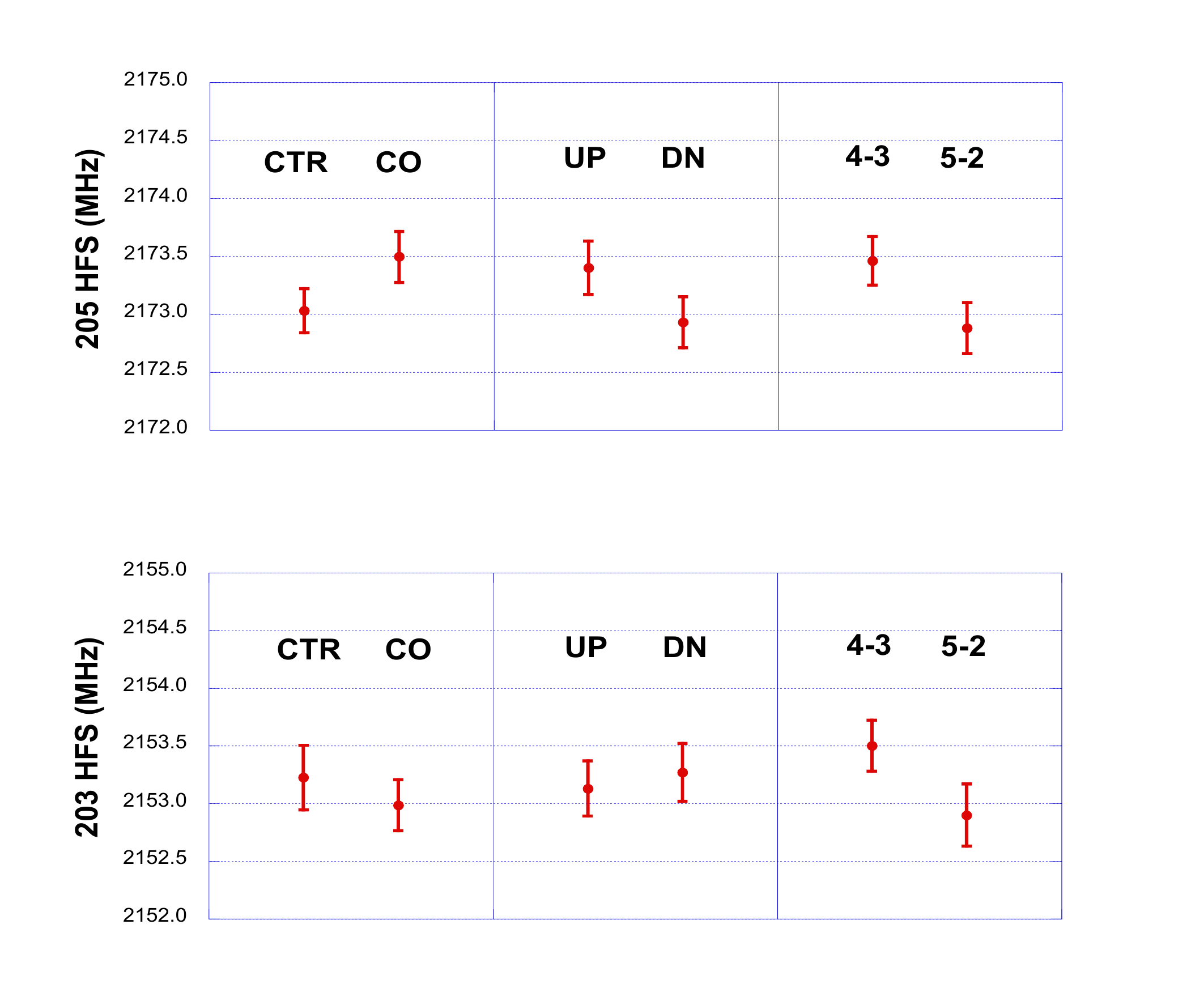}
\caption{\label{comparisons}   (Color online) Comparisons of data subsets for $^{205}$Tl (top panel) and $^{203}$Tl (bottom panel) to investigate potential systematic errors arising from  UV laser beam propagation direction, laser sweep direction, and HFS splitting derived from `sideband peaks' (4-3) vs. `carrier peaks' (5-2) .  Error bars reflect 1$\sigma$ uncertainties based on the observed data set-to-set scatter of relevant data subsets.}
\end{figure}

For all of the single-isotope data, we scanned the IR laser upward and downward in frequency, and then used the optical shutters to alternate between CO and CTR geometries.  By subdividing each data set into four categories, we can investigate possible systematic errors associated with these parameters.  Fig. \ref{comparisons} shows weighted averages of all results for each isotope when we subdivide the data in this way.  From analyses such as this we are able to place limits on systematic errors associated with laser sweep direction and beam geometry for the single-isotope results.  All such data-subset differences were well below the 1 MHz level, as noted in Fig. \ref{comparisons}.

The principal systematic error concern in this experiment is the reliability of our frequency axis calibration. In our final analysis, we used the average FM sideband-based correction factor for each day's data run to correct raw frequency splittings for that day.  Since it is conceivable that long term drifts in the FP cavity or changes in beam alignment through it could change the FP cavity FSR, we do not assume that the exact factor required to correct the nominal frequency scale \emph{ought to} remain exactly constant over time.  We find that  the sideband-based correction factor for the 205 single-isotope data set to be $\mathcal{C}_{EOM}^{205} = 1.0018(2)$ while for the 203 single-isotope data we find $\mathcal{C}_{EOM}^{203} = 1.0022(2)$.  It is reassuring that these are in statistical agreement with the less precise, but `direct' Fabry-Perot FSR measurement method, $\mathcal{C}_{FP} = 1.0024(5)$.  We also compared the calibration factors for subsets of data corresponding to different beam propagation direction and laser sweep direction. None of the calibration factors from these subsets varied from the mean calibration value by more than 0.0002.  

Next, we considered residual differences between individual sideband values (as suggested in Fig. \ref{4sidebands}), which could result from frequency scale non-linearity that persists even after Fabry-Perot spectrum analysis.  Standard deviations about the mean sideband value were typically below the 1 MHz level (out of 600 MHz).  We explored the correlation between \emph{average} value of the sideband splittings with the \emph{variance} among the four values for all of the data collected, allowing us to put tight limits on potential systematic calibration errors due to residual scan non-linearity. The excellent agreement between calibration-corrected HFS values for up scan vs. down scan results is also reassuring in this regard, since the scan non-linearity is distinctly different for each laser scan direction, due to the hysteretic nature of the PZT controlling frequency tuning.

We generate two sets of statistically independent HFS values for each single-isotope spectrum.  First, using the central HFS spectral peaks, we have: $f_{HFS}^{52} = \mathcal{C}_{EOM} * \nu_{52}$.  We also generate the quantity $f_{HFS}^{43} = \mathcal{C}_{EOM} * \nu_{43} + 1200.0$ MHz using the two `inner' first-order sidebands.  This latter HFS value, as discussed above, is somewhat less sensitive to possible errors in calibration factors.  As can be seen in Fig. \ref{comparisons}, the agreement between these HFS determinations is quite good, providing further confidence in our overall calibration procedure.  Small remaining differences between calibration values based on different methods, and for subsets of data, can be converted into equivalent frequency differences, and yields our calibration-related systematic error contribution to our final results.

For the dual-isotope scans, from which we extract the transition isotope shift (Eq. \ref{isotopeshift}), we chose to use the overall average value of  $\mathcal{C}_{EOM} = 1.0020$ to calibration-correct all scans. The relevant isotope shift is smaller by a factor of four than the HFS, so potential systematic frequency errors due to calibration inaccuracies are proportionately smaller.  Comparisons of results from upward and downward going dual-isotope spectra showed excellent statistical agreement.   The mean transition isotope shift values for all DI scans was $I_{7s-7p} = 534.4(5)$ MHz, where the error bar reflects the observed data set-to-set scatter.     We found that the HFS intervals extracted from the DI scans showed more intrinsic statistical scatter than those obtained from the single-isotope scans, but the overall mean values from the two analyses differenced by less than 1 MHz level, with errors of order 0.5 MHz (dominated by the larger error from the DI scan results).  The average value of the total Doppler shift extracted from the DI scans (Eq. \ref{Doppshift}) was roughly 0.5 MHz lower than the expected 474.7 MHz.  This small discrepancy could be attributed to slight angular misalignments of the counter-propagating beams, and would in that case result in a small systematic error in the extraction of the isotope shift, which we list at the end of Table \ref{systable}). Finally, we generated a scatter plot of the TIS value with each HFS value extracted from the same scan as well as the value of the total Doppler shift,  and searched for correlations between the measured TIS values with each of these three quantities.  This analysis produced the final systematic error entry quoted in Table \ref{systable}.  

Having calibration-corrected all frequency intervals, final mean values were computed both by taking the weighted average of all individual sweep results, as well the chi-squared-corrected average of results from full data sets.  We also fit Gaussians to the histogram of all individual HFS and IS values. Finally, we computed averages of  data subsets such as those plotted in Fig. \ref{comparisons}.  These various methods all gave final values that were in good statistical agreement with each other.  We take as our final central values for the HFS intervals the mean value of the results obtained from the single-isotope analysis of the carrier peak splitting ($\delta\nu_{52}$) and that of the sideband peaks ($\delta\nu_{43}$).

\section{Final results and discussion}
Table \ref{systable} summarizes our final results as well as final statistical errors and various sources of systematic errors.  The final combined errors for the HFS splittings make the precision of our new measurement comparable to that quoted in the 1988 measurements\cite{Grexa88}.  However, as can be seen in Table \ref{compare}, the two sets of experimental results disagree significantly.  We also list an \emph{ab initio} theoretical value of the $^{205}$Tl hyperfine splitting, computed as part of the most recent theoretical effort to compute parity non conservation (PNC) in thallium\cite{Kozlov01}.  While it is not likely that the uncertainty in the theoretical value for this quantity will ever approach the 1 MHz level, the 20 MHz discrepancy between new and old experimental values is notable, and brings the experiment and theory numbers into significantly better agreement.  Since the PNC calculation, like the hyperfine splitting calculation, focus on short-range electron wave function behavior as well as nuclear structure models, accurate experimental benchmark values for hyperfine splittings are absolutely essential for testing the accuracy of these complex calculations.

\begin{widetext}
\begin{table*}
		\caption{Summary of results and contributions to the overall 
		error in measured frequency intervals.  The final two entries in the systematic errors list pertain only to the dual-isotope analysis (see text).\label{systable}}
	\begin{tabular}{|c|r|r|r|}
		\hline\hline
	  & 
		  7p$_{1/2}$ ($^{205}$Tl) &
		  7p$_{1/2}$ ($^{203}$Tl) &
		 7s$_{1/2}$ - 7p$_{1/2}$\\
		 & HFS & HFS & Transition IS \\
		\hline\hline
		{\bf Final result (MHz)} & 2173.3 & 2153.2 & 534.4 \\
		\hline 
		&&& \\
		{\bf Statistical error (MHz)} & 0.20  & 0.25  & 0.5  \\
	\hline 
	{\bf Systematic error sources (MHz)} &&& \\

	Laser Sweep (Dir./Speed/Width) & 0.3 & 0.3 & 0.2 \\
	Beam co vs. counter-propagation & 0.3 & 0.2 & \\
	Frequency Calibration & 0.55 & 0.45 & 0.5\\
	Scan linearization & 0.2 & 0.2 & 0.2  \\
	Thallium cell temperature & 0.2 & 0.2 & 0.2 \\
	Geometrical alignment of beams & & & 0.25 \\
	Correlation with HFS, total Doppler shift & & & 0.3 \\
	\hline 
	&&& \\
		 {\bf Combined Error Total (MHz)}  &  0.8 & 0.7 & 0.9  \\
	\hline
		\end{tabular}
\end{table*}
\end{widetext}

\begin{table}
		\caption{Summary of measurements of thallium 7p$_{1/2}$-state 
		hyperfine splittings.  All results are in MHz.¥\label{compare}}
	\begin{tabular}{|cc|lc|l|}
	  	\hline
		 Source && $^{205}$Tl        &  &            $^{203}$Tl \\
		\hline\hline
		Ref. \cite{Grexa88} && 2155.5(6) && 2134.6(8) \\
		\hline 
		&&&& \\
		{\bf Present results}  && {\bf 2173.3(8)} && {\bf 2153.2(7)}  \\
		\hline
		&&&&\\
		Theory (Ref. \cite{Kozlov01}) && 2193   &  & \\
		\hline
		\end{tabular}
\end{table}

From our measured values of the hyperfine splittings, we can deduce the hyperfine anomaly,  $\Delta \equiv 
[(\mathcal{H}_{7p, 205}/\mathcal{H}_{7p, 203})(g_{203}/g_{205}) - 1]$, where the $g$'s refer to the 
nuclear g-factor of the relevant isotope.  Using very precise values for the g-factors tabulated in \cite{Raghavan87}, we 
find the barely resolved negative value: $\Delta_{7p_{1/2}} = -5(4)\times 10^{-4}$.  The sign and magnitude of this anomaly is consistent with those found for other low-lying states of thallium\cite{Chen12}. This experimental quantity can be combined with nuclear structure  
calculations regarding the magnetic moment and charge distributions in the 
isotopes to infer a value for the mean square isotopic {\em change} in these 
distributions.  Such a calculation has been performed for both the ground 6p states and the excited 7s$_{1/2}$ state\cite{ammp}, but not yet for the 7p$_{1/2}$ state.  

Regarding our transition isotope shift measurement ($^{205}$Tl  relative to $^{203}$Tl ), we note that  our result is in statistical disagreement with the original 1988 result\cite{Grexa88}, though the agreement is improved when comparing to the later paper\cite{Hermann93} which refers to a `corrected evaluation'.  Using the value derived for the level isotope shift of the 7s$_{1/2}$ state of +409.0(3.8) MHz\cite{Richardson00}, we can infer a 7p$_{1/2}$-state level isotope shift  of $I_{7p} = -125.4(4.0)$ MHz.

\section{Concluding remarks}
Using two-step, two-color diode laser spectroscopy, we have measured the 7p$_{1/2}$-level hyperfine splittings in $^{205}$Tl and $^{203}$Tl, as well as the transition isotope shift in the 1301 nm $7s_{1/2} - 7p_{1/2}$ transition.  The HFS values are each roughly 20 MHz larger than previously published values, and show improved agreement with theoretical estimates.  Rather than pursue greater precision in our current measurement, we believe it is critically important to re-measure other hyperfine intervals, especially given that systematic and calibration errors may persist in the literature.  The importance of this element in atomic-physics-based tests of  discrete symmetry violation, which require precise, independent atomic theory calculations, helps to motivate this experimental effort.  

As a straightforward next step, we are presently installing a red diode laser (wavelength near 670 nm)  to substitute for the the IR laser in our thallium vapor cell setup.  With our UV laser again locked to the ground-state 377.6 nm transition, we will then probe the $7s_{1/2} - 8p_{1/2, 3/2}$ transitions.  Using an identical spectroscopy technique and FM-sideband calibration scheme, it should be straightforward to measure these hyperfine splittings with uncertainties below 1 MHz, allowing us to again compare new results to theory, and to previously measured values for these hyperfine intervals.

\begin{acknowledgments}
We would like to thank Taryn Siegel for her work at the initial stages of this experiment and Michael Taylor for his expert advice  and aid in mechanical design.  We gratefully acknowledge the support of the National Science Foundation RUI program, through grant No. 0969781.
\end{acknowledgments}

\bibliography{Tlbiblio}

\end{document}